\documentclass[aps,prl,twocolumn,showpacs,superscriptaddress,groupedaddress]{revtex4}  
\usepackage{graphicx}  
\usepackage{dcolumn}   
\usepackage{bm}        
\usepackage{amssymb}   
\usepackage{color}
\usepackage{pdfpages}

\begin{document}

\title{Non-reciprocal mu-near-zero mode in $\mathcal{PT}$-symmetric magnetic domains}

\author{Jin Wang} \affiliation{Department of Physics, Southeast University, Nanjing 211189, China}\affiliation{Department of Applied Physics, The Hong Kong Polytechnic University, Hong Kong, China}
\author{Hui Yuan Dong} \affiliation{School of Science, Nanjing University of Posts and Telecommunications, Nanjing 210003, China}
\author{Chi Wai Ling} \affiliation{Department of Applied Physics, The Hong Kong Polytechnic University, Hong Kong, China}
\author{C.~T.~Chan}  \affiliation{Department of Physics, Hong Kong University of Science and Technology, Clear Water Bay, Hong Kong, China}
\author{Kin Hung Fung} \email{khfung@polyu.edu.hk}\affiliation{Department of Applied Physics, The Hong Kong Polytechnic University, Hong Kong, China}
\date{\today}

\begin{abstract}

We find that a new type of non-reciprocal modes exists at an interface between two \emph{parity-time} ($\mathcal{PT}$) symmetric magnetic domains (MDs) near the frequency of zero effective permeability. The new mode is non-propagating and purely magnetic when the two MDs are semi-infinite while it becomes propagating in the finite case. In particular, two pronounced nonreciprocal responses could be observed via the excitation of this mode: one-way optical tunneling for oblique incidence and unidirectional beam shift at normal incidence. When the two MDs system becomes finite in size, it is found that perfect-transmission mode could be achieved if $\mathcal{PT}$-symmetry is maintained. The unique properties of such an unusual mode are investigated by analytical modal calculation as well as numerical simulations. The results suggest a new approach to the design of compact optical isolator.

\end{abstract}

\maketitle

\section{\label{sec:level1}I. INTRODUCTION}

Over the past few decades there has been much activity on the non-reciprocity effect in optics\cite{Kong,Figotin,Haldane,Raghu,Wang,Poo,Yu,Lira,Alu,Manipatruni,Kang,Gallo,Soljacic,Fan,Xu,Dong,Zhu,Zhang}. Non-reciprocal optical elements, such as optical isolators, have attracted great attention owing to its capability of allowing light to propagate only along a single direction, while strongly suppressing backward scattering. The traditional way for creating nonreciprocal devices relies on magneto-optic Faraday effect in the presence of an external magnetic field. However, the intrinsic weakness of Faraday effects based on available magneto-optical (MO) materials makes the Faraday rotator bulky and hinders miniaturization of such devices. Later, the photonic crystal (PC) made of MO materials \cite{Figotin} was suggested to enhance the nonreciprocal response, and create compact and integrated isolators and circulators. Recently, Raghu and Haldane \cite{Haldane,Raghu} theoretically predicted one-way edge modes could be observed in MO photonic crystals, as optical counterparts to chiral edge states of electrons in the quantum Hall effect. These modes are confined to the region near the edge of the 2D PC, displaying one-way propagation characteristics. Subsequently, experimental realizations and observations of such electromagnetic one-way edge states in different magneto-optical photonic crystal (MPCs) were reported by several groups \cite{Wang,Poo}. Nonreciprocal behavior has also been demonstrated by considering dynamic modulation in standard materials \cite{Yu,Lira,Alu}, the use of opto-mechanical \cite{Manipatruni} and opto-acoustic effects \cite{Kang} and optical nonlinearities \cite{Gallo,Soljacic,Fan,Xu}.

On the other hand, considerable efforts have been intensively devoted to a new class of artificial optical materials having balanced loss and gain - \emph{parity-time} ($\mathcal{PT}$)-symmetric metamaterials \cite{Bender,Makris,Regensburger,Lin,Zhu2,Feng,Ramezani,Ruter,Peng,Nazari,Hernandez-Coronado,Jiri,Maziar,Ge,Ge2,Savoia}. Such $\mathcal{PT}$-symmetric systems have non-Hermitian Hamiltonians, exhibiting with entirely real eigenvalues as long as $\mathcal{PT}$ symmetry holds. Remarkably, the system may undergo an abrupt phase transition (spontaneous $\mathcal{PT}$ symmetry-breaking) at some non-Hermiticity threshold, beyond which some of the eigenvalues become complex. To date, several $\mathcal{PT}$-symmetric models have been demonstrated with some intriguing light propagation behaviors, including power oscillations \cite{Makris}, double refraction \cite{Makris}, unidirectional invisibility \cite{Regensburger,Lin,Zhu2,Feng}, non-reciprocal light transmission \cite{Ramezani,Ruter,Peng,Nazari} and unattenuated surface modes \cite{Savoia, Jiri, Maziar}.

It turns out that $\mathcal{PT}$-symmetry has a strong linkage to perfect transmission states \cite{Hernandez-Coronado}. This type of spatial-temporal symmetry can be more general than the usual symmetry-related perfect transmission associated with mirror symmetry or inversion symmetry. Since such a $\mathcal{PT}$-symmetry-related perfect transmission is complementary to non-reciprocity, it is also useful for the design of optical isolator displaying one-way perfect transmission with no gain medium such as the case in this paper. In the present work, we consider a structure composed of two MDs with $\mathcal{PT}$ symmetry \cite{Zhu,Zhang}, magnetized homogenously in opposite directions, and find a new type of non-reciprocal mu-near-zero (MNZ) modes at the interface separating two MDs near the frequency of zero effective permeability. The broken $\mathcal{P}$ and $\mathcal{T}$ symmetries, induced here simultaneously by the geometry and the orientation of the external magnetic field, result in the asymmetrical dispersion relations of the interface mode, whereas the unbroken $\mathcal{PT}$ symmetry leads to the emergence of the perfect transmission mode \cite{Hernandez-Coronado}. Furthermore, two pronounced nonreciprocal behaviors are exhibited by application of such a MNZ mode for incident plane waves: one-way complete optical tunneling at oblique incidence and unidirectional beam shift at normal incidence. Calculations on nonreciprocal dispersion relations, reflection spectra and field patterns for such a $\mathcal{PT}$-symmetric system are employed to verify our conclusions.

This paper is organized as follows. In Sec. II, the exact analytical modal description is employed to investigate the non-reciprocal MNZ mode in the $\mathcal{PT}$-symmetric system we proposed. Sec. III shows the numerical results of reflection spectra and field patterns for the finite-size $\mathcal{PT}$-symmetric system. Finally, the conclusions are given in Sec. IV.

\section{II. ANALYTICAL MODAL DESCRIPTION OF NON-RECIPROCAL Mu-Near-Zero MODE}

We start with two semi-infinite MDs constructed by MO media oppositely magnetized in the Voigt geometry as shown in Fig. 1(a). Under the external static magnetic field along $\pm z$, the two semi-infinite MDs are, characterized respectively by identical permittivities $\epsilon_{m}$ and magnetic permeability tensors $\bar{\mu}_{(x>0)}$ and $\bar{\mu}_{(x<0)}$ \cite{Zhu,Zhang},
\begin{equation}
\bar{\mu}_{(x>0)}=
\left(
\begin{tabular}{ccc}
$\mu_{1}$ & $i\Delta_{1}$ & $0$\\
$-i\Delta_{1}$ & $\mu_{1}$ & $0$\\
$0$ & $0$ & $\mu_{1}$\\
\end{tabular}
\right),
\bar{\mu}_{(x<0)}=
\left(
\begin{tabular}{ccc}
$\mu_{2}$ & $i\Delta_{2}$ & $0$\\
$-i\Delta_{2}$ & $\mu_{2}$ & $0$\\
$0$ & $0$ & $\mu_{2}$\\
\end{tabular}
\right).
\end{equation}
We take the following parameters for MDs \cite{Poo}, i.e., $\mu_{1}=\mu_{2}=1+\omega_{m}\omega_{h}/(\omega_{h}^{2}-\omega^{2})$, $
\Delta_{1}=-\Delta_{2}=-\omega_{m}\omega/(\omega_{h}^{2}-\omega^{2})$, where $\omega_{h}=\gamma H_{0}$ is the precession frequency, $\gamma$ is the gyromagnetic ratio, $H_{0}$ is the applied magnetic field on the two MDs, $\omega_{m}=4\pi\gamma M_{s}$, and $4\pi M_{s}$ is the saturation magnetization. The parameters are chosen to fulfill $\mathcal{PT}$ symmetry $\bar{\mu}_{(x>0)}=\bar{\mu}^{*}_{(x<0)}$, which will lead to perfect transmission modes. The complex conjugate in $\bar{\mu}$ is associated with time-reversal operation (see Appendix A). It should be noted that only transverse electric (TE) polarization (i.e., electric field along the $z$ direction) is considered, and the $e^{-i\omega t}$ time-dependent convention for harmonic field is used in this work.

\begin{figure}[!htbp]
\includegraphics[width=3.2in]{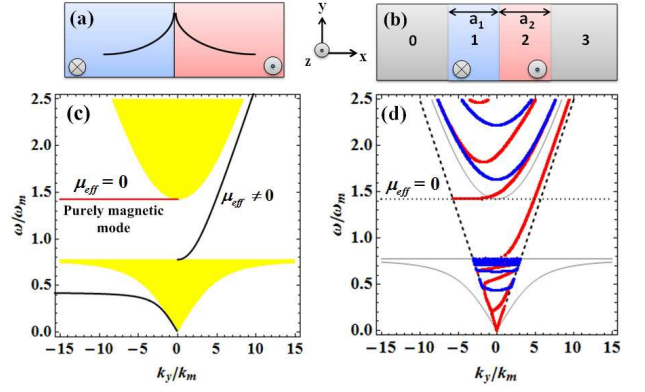}
\centering
\caption {\small (color online) (a),(b) Schematic diagram of the $\mathcal{PT}$-symmetric system. (a) Two semi-infinite MDs locate at $x>0$ and $x<0$, respectively. (b) Finite-size bilayer MD slabs composed of two halves of identical thickness $a_1=a_2=a$, embedded in surrounding mediums (with refractive index $n$). Labels $0,1,2,3$ are used to indicate four different regions in our system. (c) The dispersion relation of interface modes in (a). Yellow and white regions represent bands and gaps of an infinite MD, respectively. (d) The radiative modes in (b). The bulk band-edge for magnetic materials (gray lines), light curves for surround mediums (dashed lines), and the frequency line (dotted lines) corresponding to zero effective permeability of MD are also shown. In (c)(d), we set $k_m=\omega_{m}/c$ (here $c$ denotes the speed of light in vacuum) as a scale to represent the transversal wave vector $k_y$. For the mode solutions in (d) for the finite-size bilayer MDs structure, each magnetic layer is assumed to have equal thickness $a=0.008$ m, and the surrounding medium with refractive index $n=4$.}
\end{figure}

Before we solve for the solutions of the interface modes, it should be noted that each MD also supports bulk modes given by the dispersion relation, $k^{2}=\epsilon_{m}\mu_{eff}\omega^{2}/c^{2}$, where $\mu_{eff}$ is the effective permeability defined as $\mu_{eff}=(\mu_{1}^{2}-\Delta_{1}^{2})/\mu_{1}=(\mu_{2}^{2}-\Delta_{2}^{2})/\mu_{2}$ and $\textbf{k}=(k_{x},k_{y},0)$ is the wavevector in the $xy$-plane. Due to the resonance feature of $\mu_{1}(\mu_{2})$, a typical resonance gap is opened and the bulk modes are divided into two groups of bands for $\mu_{eff}>0$ as shown in Fig. 1(c), with the upper bands bounded by $\omega>\omega_{0}(=\omega_{h}+\omega_{m})$ and $k_{y}^{2}<\epsilon_{m}\mu_{eff}{\omega^2}/{c^2}$, and the lower bands bounded by $\omega<\sqrt{\omega_{h}(\omega_{h}+\omega_{m})}$ and $k_{y}^{2}<\epsilon_{m}\mu_{eff}{\omega^2}/{c^2}$.

To form guided waves at the interface between two MDs, the field should decay exponentially away from the interface, and can be written as follows: $\textbf{E}(x>0)=(0,0,A)e^{-\alpha x+i k_{y} y}$ and $\textbf{E}(x<0)=(0,0,B)e^{\beta x+i k_{y} y}$. Here, $A$ and $B$ are the amplitudes of the corresponding electric field components in two MDs. $\alpha$ and $\beta$ denote positive decay parameters, displaying the relations with the parallel component of wave vector $k_y$: $k_{y}^{2}-\alpha^{2}=k_{y}^{2}-\beta^{2}=\epsilon_{m}\mu_{eff}\omega^{2}/c^{2}$ in two homogenous gyromagnetic materials, with identical effective permeability $\mu_{eff}$. By solving the Maxwell's equations, we have magnetic fields components $\textbf{H}=(H_{x},H_{y},0)e^{-\alpha x+i k_{y} y}$ for the $x>0$ space satisfying the following relations:
\begin{eqnarray}
(\mu_{1}^2-\Delta_{1}^2) H_{x}(x>0)&=&\frac{A}{\omega}(\mu_{1} k_{y}-\Delta_{1}\alpha),\nonumber\\
(\mu_{1}^2-\Delta_{1}^2) H_{y}(x>0)&=&-i \frac{A}{\omega}(\mu_{1}\alpha -\Delta_{1}k_{y}). \label{magfld}
\end{eqnarray}
By replacing $A$, $\alpha$ and $\Delta_{1}$ by $B$, $-\beta$ and $-\Delta_{2}$, respectively, we could obtain the corresponding equations of magnetic field for the space $x<0$.

In most cases that the condition $\mu_{eff}\neq 0$ is fulfilled, the magnetic field could be then easily obtained from Eq.(\ref{magfld}). With the boundary condition that the tangential field components should be continuous across the interface, we could have the usual ``$\mu_{eff}\neq 0$" solution for an interface mode, shown with black solid lines in Fig. 1(c) as well as in Ref.\cite{Zhang}. More interestingly, if we take into account the possibility of $\mu_{eff}=0$ (here $\mu_{1}=\Delta_{1}$) at $\omega_{0}=\omega_{h}+\omega_{m}$ in this specified case, there exists an extra solution of interface mode in this $\mathcal{PT}$-symmetric system:
\begin{equation}
A=B=0,~\alpha=\beta=-k_{y}. \label{disp}
\end{equation}
We called such a non-trivial solution the $\mu_{eff}=0$ mode. It is interesting that the mode is purely magnetic with no electric field while the two orthogonal components of magnetic field has the following unique relations:
\begin{equation}
H_{x}(x>0)=-H_{x}(x<0)=-i H_{y}, \label{magnetic}
\end{equation}
indicating the certain phase difference between $H_x$ and $H_y$ with $\pi/2$ in the left domains region and $-\pi/2$ at the right. Moreover, in order to guarantee the positive decay rate ($\beta>0$, $\alpha>0$), the parallel component of wave vector $k_y$ should remain negative, which leads to the emergence of a nonreciprocal $\mu_{eff}=0$ mode shown by the red line in Fig. 1(c). Here, we use parameters for MDs provided in a previous experimental study \cite{Poo}, i.e. $\epsilon_{m}=15.26, H_{0}=800$ Oe, and $4 \pi Ms=1884$ G.

However, the nonreciprocal $\mu_{eff}=0$ modes between two semi-infinite domains form a flat band and thus they are non-propagating, which makes the modes difficult to be excited. To improve its optical response, we alter the infinite systems by the finite-size bilayer MDs still with $\mathcal{PT}$ symmetry [shown in Fig. 1(b), and here assumed with identical thickness $a_{1}=a_{2}=a$], embedded in an uniform surrounding medium. Based on the transfer matrix approach \cite{Dong}, the radiative modes for such a bilayer system outside the light line for surrounding mediums could be well solved. Two kinds of mode solutions could be analytically separated as
\begin{equation}
\frac{\sin (k_{x} a)}{k_x}=0 \label{symm}
\end{equation}
for reciprocal (symmetrical) modes and
\begin{eqnarray}
&&\frac{1}{k_{x0}}[ \cos (k_{x} a) (k_{y}^{2}(\frac{\mu_{0}}{\mu_{1}}-\frac{\mu_{eff}}{\mu_{0}})-\frac{\omega^{2}}{c^{2}}(\epsilon_{m}\mu_{0}-\epsilon_{0}\mu_{eff})) \nonumber \\
&+&\frac{k_y \Delta_{1}}{k_x \mu_1}\sin(k_{x} a) (k_{y}^{2}(\frac{\mu_{0}}{\mu_{1}}+ \frac{\mu_{eff}}{\mu_{0}})-\frac{\omega^{2}}{c^{2}}(\epsilon_{m}\mu_{0}+\epsilon_{0}\mu_{eff}))
]   \nonumber \\
&=&0 \label{asymm}
\end{eqnarray}
for non-reciprocal (asymmetrical) ones (Appendix B gives the derivation of Eq. (\ref{symm}) and (\ref{asymm})). Here, $\epsilon_{0}$ and $\mu_{0}$ are the permittivity and permeability for surrounding medium, and the wave-vector components normal to the interface in background and magnetic materials are taken as $k_{x0}=\sqrt{\epsilon_{0}\mu_{0}\omega^{2}/c^{2}-k_{y}^{2}}$, and $k_{x}=k_{x1}=k_{x2}=\sqrt{\epsilon_{m}\mu_{eff}\omega^{2}/c^{2}-k_{y}^{2}}$, respectively. The reciprocal propagating modes in Eq.(\ref{symm}) for such bilayer MD systems are identical to those in a single slab layer of MD, simultaneously independent of surrounding mediums.  It should be emphasized that the linear term of $k_y$ in Eq.(\ref{asymm}) breaks the spectral reciprocity (i.e., the left-right symmetry of the dispersion relation), leading to strong non-reciprocal behaviors. Furthermore, in the limit of $a\rightarrow\infty$, there is always a solution at $\omega_{0}$ identical with Eq.(\ref{disp}) for the infinite system in Fig. 1(a).

We plot in Fig. 1(d) the corresponding radiative electromagnetic modes within the light cone for surrounding media with the refractive index $n=4$. Each magnetic layer has equal thickness $a=0.008$ m. The reciprocal and non-reciprocal modes are shown by blue and red lines, respectively. It is found that the original flat and non-propagating $\mu_{eff}=0$ mode interacts with the propagating modes in bilayer MDs, and extends to the bulk band for magnetic materials, thereby becoming dispersive. So we achieve a non-reciprocal mu-near-zero ($\mu_{eff}\simeq 0$) radiative mode for thin films of MD structures, and expect to see the non-reciprocal optical response for the dispersive mode near the frequency $\omega_{0}$ corresponding to $\mu_{eff}=0$, with direct illumination of external plane waves.

\begin{figure}[!htbp]
\includegraphics[width=3.2in]{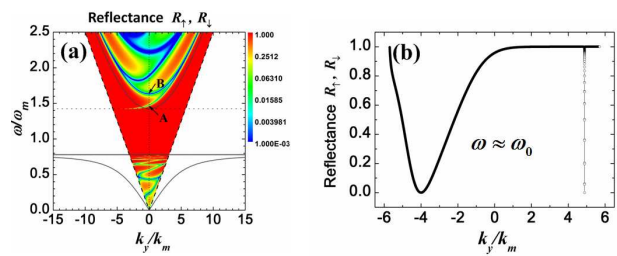}
\centering
\caption {\small (color online) (a)(b) The reflectance spectra for finite-size $\mathcal{PT}$-symmetric MDs shown in Fig. 1(b), where $R_{\uparrow}$ and $R_{\downarrow}$ represent, respectively, the reflectances for upward $(k_{y}>0)$ and downward $(k_{y}<0)$ rays either incident from left or right. (a) Contour plot - reflectance as a function of $\omega$ and $k_y$, (b) 2D line plot - reflectance as a function of $k_y$ with a frequency $\omega=1.425\omega_{m}$ close to $\omega_{0}$, just along the horizontal dotted line in (a). Gray and dashed lines in (a), and the MDs structure parameters used here are the same as those in Fig. 1(d).}
\end{figure}

\section{III. NUMERICAL RESULTS ON FINITE-SIZE $\mathcal{PT}$-SYMMETRIC MAGNETIC DOMAINS}

To support our findings, we investigate the wave propagation behaviors through finite-size $\mathcal{PT}$-symmetric MDs, with numerical calculations on the reflection spectra [shown in Fig. 2], where $R_{\uparrow}$ and $R_{\downarrow}$ represent, respectively, the reflectances for upward $(k_{y}>0)$ and downward $(k_{y}<0)$ rays either incident from left or right. Apparently, it is seen that reflectance dips shown as dark blue colors in Fig. 2(a) are in excellent agreement with those radiative modes in Fig. 1(d), and the dispersive and non-reciprocal $\mu_{eff}\simeq 0$ mode could be well excited under external plane waves, as shown in Fig. 2(b) with a particular example of the frequency $\omega=1.425\omega_{m}$ close to $\omega_{0}$. In contrast to the usual $\mu_{eff}\neq 0$ interface mode indicated with a very narrow dip in Fig. 2(b), the coupled $\mu_{eff}\simeq 0$ mode shows strong non-reciprocity response over a much wider region of the incident angle.

It should be noted that in a one-dimensional $\mathcal{PT}$-symmetric system with balanced gain and loss, there exists a new conservation rule $|1-T|=\sqrt{R_{L}R_{R}}$ \cite{Ge}, where $T$ is the transmittance through the entire system, $R_{L}$ and $R_{R}$ are, respectively, the reflectances for left and the right rays traveling either upwards or downwards. Such a system is reciprocal in the linear regime. In contrast, the non-reciprocal ``Hermitian" system discussed in this paper obeys the standard conservation laws $1-T_{\uparrow}=R_{\uparrow}$ and $1-T_{\downarrow}=R_{\downarrow}$ instead for upward and downward rays, even the transmittances in opposite directions ($T_{\uparrow}$ and $T_{\downarrow}$) are different (See Appendix C for discussion on scattering problems in a multi-port system). Nevertheless, our system is still a $\mathcal{PT}$-symmetric system without single $\mathcal{P}$ or $\mathcal{T}$ symmetry.

\begin{figure}[!htbp]
\includegraphics[width=3.2in]{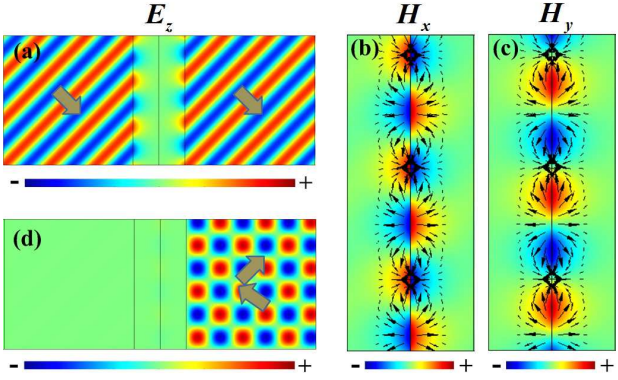}
\centering
\caption {\small (color online) Electric-field distribution at $\omega=1.425\omega_{m}$ under front illumination with $k_{y}=-4 k_{m}$ (a), and back illumination with $k_{y}=4 k_{m}$ (d). Magnetic-field patterns in (a) of $H_{x}$ (b), and $H_{y}$ (c) in the regions filled with the two gyromagnetic materials at a zoom-in view. Black arrows in (b)(c) show vector patterns of the magnetic field $\textbf{H}=(H_{x}, H_{y})$. The MDs structure parameters are the same as those used in Fig. 1(d). The big arrows shown in (a) and (d) guide us to see the wave propagation.}
\end{figure}

Further, 2D finite-element simulations using COMSOL Multiphysics were carried out to verify the electromagnetic non-reciprocal response of waves impinging on our proposed finite-size $\mathcal{PT}$-symmetric systems. Figure 3 depicts the spatial field distribution with a frequency of $\omega=1.425\omega_{m}$ at oblique incidence. Counter-propagating plane waves are incident from surrounding mediums upon either side of the bilayer MD structures. For the case of the downward incidence shown in Fig. 3(a), full transmission could be obtained due to the excitation of $\mu_{eff}\simeq 0$ mode on the interface. Interestingly, it is found that there exists a purely magnetic field with no electric field along the interface. To see more clearly, we zoom in and get a close-up view of the magnetic field $\textbf{H}=(H_{x}, H_{y})$ in the two domains as shown in Fig. 3(b)(c), with black arrows representing the vector patterns of magnetic field.  The fixed phase difference between $H_x$ and $H_y$ could be observed, such as $\pi/2$ in the left domain region and $-\pi/2$ at the right. These results are identical with the derivation of Eq.(\ref{magnetic}) for the infinite system. In contrast, for upward incidence in Fig. 3(d), such excitation of $\mu_{eff}\simeq 0$ mode is almost completely suppressed, resulting in low transmission through the structure. Therefore a non-reciprocal optical response is attained with one-way tunneling for incident oblique waves through thin films of $\mathcal{PT}$-symmetric bilayer MD structure.

\begin{figure}[!htbp]
\includegraphics[width=3.2in]{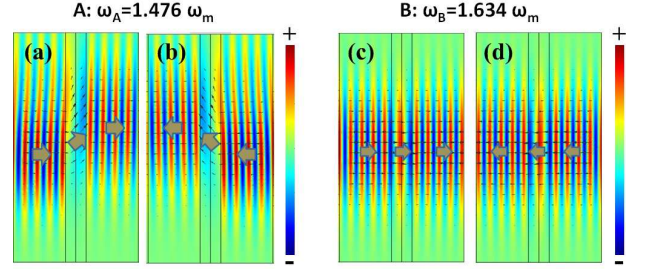}
\centering
\caption {\small (color online) Electric field distribution under the (a) front illumination and (b) back illumination of an incident gaussian wave normal to interface with a frequency of $\omega_{A}=1.476 \omega_{m}$. (c)(d) are similar to (a)(b), but for another case with a frequency $\omega_{B}=1.634 \omega_{m}$. These two particular cases are marked in Fig. 2(a) with point A and B, respectively. The vector patterns of power flow in our system are also illustrated with black arrows in (a)-(d). The MDs structure parameters are the same as those used in Fig. 1(d). The profile of incident Gaussian beam is assumed to be $|E_{z}|=E_{0}\exp{(-y^{2}/0.002)}$ (SI unit), where $E_{0}$ determines the arbitrary overall amplitude in the linear regime. For clear illustration, the power flow of incident waves is also shown by means of big arrows.}
\end{figure}

At normal incidence shown in Fig. 4, another interesting phenomenon of non-reciprocal beam shift could be seen by application of the $\mu_{eff}\simeq 0$ mode through such a finite $\mathcal{PT}$-symmetric structure. In Fig. 4(a)(b) at a frequency of $\omega_{A}=1.476 \omega_{m}$ [corresponding to point A shown in Fig. 2(a)], both incoming Gaussian waves, including from left or right, undergo an upward lateral-shift perpendicular to the propagation direction after passing through the bilayer MDs. Meanwhile, by looking inside the magnetic domains at both of incidence cases, the direction of power flow indicated by black arrows always changes by an upswept angle with respect to the power flow of the incoming waves. The beam shift and non-reciprocal behavior can also be understood by the excitation of $\mu_{eff}\simeq 0$ mode at point A, with an upswept-angle direction of wave group velocity $\mathbf{v}_{g}$, evaluated as $\mathbf{v}_{g}=\vec{\nabla}_{\mathbf{k}}\omega(\mathbf{k})$ from the dispersion relation of Fig. 1(d). For comparison, at another resonant frequency of $\omega_{B}=1.634 \omega_{m}$ [corresponding to point B in Fig. 2(a)], the incoming waves go straightforward with reciprocal response shown in Fig. 4(c)(d), because the reciprocal propagating mode is excited with the group velocity at point B keeping along the horizontal direction.

We emphasis that the $\mathcal{PT}$ symmetry in our system is actually not a necessary condition to achieve the spectral non-reciprocity. Nevertheless, the $\mathcal{PT}$ symmetry can help achieving perfect transmission mode in one direction as depicted in Fig. 5(a). For a non-$\mathcal{PT}$-symmetric structure with different applied magnetic field on the two magnetic domains, it is seen that transmission through the entire system would be partly suppressed, and the $\mu_{eff}\simeq 0$ mode shift slightly.

Finally, owing to the possible difficulty in implementation in practice of our proposed finite-size $\mathcal{PT}$-symmetric structures, with two adjoined, but inversely magnetized MDs, we consider another structure by separating these two MDs with a little displacement, as illustrated in Fig. 5(b). Note that the $\mu_{eff}\simeq 0$ mode shifts to the lower frequency shown in Fig. 5(c), due to the variation of the effective index of the structures.

\begin{figure}[!htbp]
\includegraphics[width=3.2in]{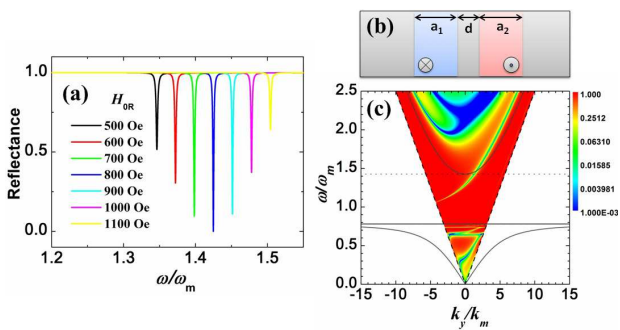}
\centering
\caption {\small (color online) (a) The reflectance spectra at a specified incident angle with $k_{y}=-4 k_{m}$ for a non-$\mathcal{PT}$-symmetric bilayer domain structure, with different applied magnetic field $H_{0R}$ on the right domain $(0<x<a)$. Here the applied field on the left domain $(-a<x<0)$ is fixed with $H_{0L}=800$ Oe, and other parameters are the same with those in Fig. 1(d). (b) Schematic diagram of two bilayer MDs, similar to Fig. 1(b), but separating them with a little displacement of horizontal distance $d$. (c) The reflectance spectra for the structure in (b), with $a_{1}=a_{2}=0.004$ m, and $d=0.002$ m. Other parameters and lines are identical with those in Fig. 1(d).}
\end{figure}

\section{IV. CONCLUSION}

In summary, we demonstrate a new type of non-reciprocal mu-near-zero radiative mode in the $\mathcal{PT}$-symmetric bilayer MDs, magnetized by opposite directions. Such an unusual mode occurs close to the frequency when the effective permeability for MDs approaches to zero, and could be well excited when the infinite system shrinks to a finite one. In particular, we see two pronounced non-reciprocal behavior for incident waves: one-way complete optical tunneling for oblique incident waves and unidirectional beam shift for normal incidence. Our theoretical results may provide a new way for designing compact isolators.

\section{ACKNOWLEDGMENTS}

This work was supported in part by the National Science Foundation of China under Grant No. 11204036, the Hong Kong Research Grant Council through the Area of Excellence Scheme (grant no. AoE/P-02/12), and the Hong Kong Polytechnic University through grant no. G-UA95.

\section{APPENDIX A: Time-reversal Symmetry}
The $\mathcal{PT}$ symmetry condition $\bar{\mu}_{(x>0)}=\bar{\mu}^{*}_{(x<0)}$ for our system has a complex conjugate on permeability tensor, which is associated with $\mathcal{T}$ operation. We note that our arguments on time-reversal symmetry are based on the following assumptions:

I. The Maxwell's equations themselves are maintained under time reversal of vector fields. The pseudo-vectors must be modified accordingly (i.e., a change in sign) in order to keep the Maxwell's equation unchanged under time-reversal.

II.	The constitutive relations among the fields (satisfying the Maxwell's equations) in frequency domain may not be the same after time reversal. Therefore, some systems are not time-reversal symmetric.

\subsection{Part I: Change in signs of pseudo-vectors}

This part is only about the change in sign related to the Maxwell's equation (not the constitutive relations).

Assume that we have the four fields (\textbf{E}, \textbf{D}, \textbf{B}, \textbf{H}) satisfying the Maxwell's equations:
\begin{eqnarray}
\nabla\times \textbf{E}&=&-\frac{\partial \textbf{B}}{\partial t},\\
\nabla\times \textbf{H}&=&\frac{\partial \textbf{D}}{\partial t}.
\end{eqnarray}
Here, we consider the solutions in source-free regions and check the conditions on the pseudo-vectors $\textbf{B}$ and $\textbf{H}$ to ensure that the equations are maintained under time reversal of vector fields $\textbf{E}$ and $\textbf{D}$.

We denote all the fields after this time-reversal operation as $\textbf{E}'$, $\textbf{D}'$, $\textbf{B}'$, $\textbf{H}'$, where we already know that $\textbf{E}'(t)=\textbf{E}(-t)$ and $\textbf{D}'(t)=\textbf{D}(-t)$ and require that the Maxwell's equations must be maintained:
\begin{eqnarray}
\nabla\times \textbf{E}'&=&-\frac{\partial \textbf{B}'}{\partial t},\\
\nabla\times \textbf{H}'&=&\frac{\partial \textbf{D}'}{\partial t}.
\end{eqnarray}
One can check that the above equations can be satisfied by the substitutions of $\textbf{B}'(t)=-\textbf{B}(-t)$ and $\textbf{H}'(t)=-\textbf{H}(-t)$ (as shown below):
\begin{eqnarray}
\nabla\times \textbf{E}'(t)&=&\nabla\times \textbf{E}(-t)=-\frac{\partial \textbf{B}(-t)}{\partial (-t)}=-\frac{\partial \textbf{B}'(t)}{\partial t} \nonumber \\
\nabla\times \textbf{H}'(t)&=&-\nabla\times \textbf{H}(-t)=-\frac{\partial \textbf{D}(-t)}{\partial (-t)}=\frac{\partial \textbf{D}'(t)}{\partial t} \nonumber
\end{eqnarray}
This means that the change in sign of pseudo-vectors is associated with the Maxwell's equations. The above results are not new and well documented in the literature \cite{Fushchich}.

\subsection{Part II: Complex conjugate in frequency domain}

We now consider the constitutive relations in frequency domain using the conclusion in Part I. We have the original four fields satisfying the following equations:
\begin{eqnarray}
\textbf{D}(\omega)&=&\bar{\epsilon}(\omega)\cdot\textbf{E}(\omega),\\
\textbf{B}(\omega)&=&\bar{\mu}(\omega)\cdot\textbf{H}(\omega).
\end{eqnarray}
It is well known that an additional complex conjugate must be applied to the frequency-domain fields when time is reversed.
Substituting $t'=-t$ into $\textbf{D}(t')=\textnormal{Re}\left(\int_{-\infty}^{\infty}\textbf{D}(\omega)e^{-i\omega t'}d\omega\right)$ will give
\begin{eqnarray}
\textbf{D}(t')&=&\textbf{D}(-t) \nonumber \\
&=&\textnormal{Re}\left(\int_{-\infty}^{\infty}\textbf{D}(\omega)e^{i\omega t}d\omega\right) \nonumber \\
&=&\textnormal{Re}\left(\int_{-\infty}^{\infty}[\textbf{D}(\omega)e^{i\omega t}]^{*}d\omega\right) \nonumber \\
&=&\textnormal{Re}\left(\int_{-\infty}^{\infty}\textbf{D}^{*}(\omega)e^{-i\omega t}d\omega\right), \nonumber
\end{eqnarray}
which gives $\textbf{D}'(\omega)=\textbf{D}^{*}(\omega)$.
Together with the conclusion in Part I, the fields in frequency domain are
\begin{eqnarray}
\textbf{E}'(\omega)&=&\textbf{E}^{*}(\omega), \textbf{D}'(\omega)=\textbf{D}^{*}(\omega), \label{TRfield1}\\
\textbf{B}'(\omega)&=&-\textbf{B}^{*}(\omega), \textbf{H}'(\omega)=-\textbf{H}^{*}(\omega). \label{TRfield2}
\end{eqnarray}
If the system is the same under time reversal, one must have
\begin{eqnarray}
\textbf{D}'(\omega)&=&\bar{\epsilon}(\omega)\cdot\textbf{E}'(\omega),\\
\textbf{B}'(\omega)&=&\bar{\mu}(\omega)\cdot\textbf{H}'(\omega).
\end{eqnarray}
The above equations are satisfied by all time-reversed fields in Eqs. (\ref{TRfield1}) and (\ref{TRfield2}) if $\bar{\epsilon}^{*}(\omega)=\bar{\epsilon}(\omega)$ and $\bar{\mu}^{*}(\omega)=\bar{\mu}(\omega)$.

Finally, we conclude that if we consider the change in sign for pseudo-vectors, the way to break time-reversal symmetry is to make either $\bar{\epsilon}^{*}(\omega)\neq\bar{\epsilon}(\omega)$ or $\bar{\mu}^{*}(\omega)\neq\bar{\mu}(\omega)$.

\section{APPENDIX B: Derivation of Equations (\ref{symm}) and (\ref{asymm})}
We start with the 1D transfer matrix $\hat{T}$ from region $0$ to $3$ [shown in Fig. 1(b)], defined by
\begin{equation}
\left(
\begin{tabular}{ccc}
$E_{3}^{+}$ \\
$E_{3}^{-}$
\end{tabular}
\right)
=\hat{T}
\left(
\begin{tabular}{ccc}
$E_{0}^{+}$ \\
$E_{0}^{-}$
\end{tabular}
\right)
=
\left(
\begin{tabular}{ccc}
$T_{11}$ & $T_{12}$ \\
$T_{21}$ & $T_{22}$
\end{tabular}
\right)
\left(
\begin{tabular}{ccc}
$E_{0}^{+}$ \\
$E_{0}^{-}$
\end{tabular}
\right).
\end{equation}
Here, $\hat{T}=\hat{M}_{23}\hat{P}_{2}\hat{M}_{12}\hat{P}_{1}\hat{M}_{01}$ is the total transfer matrix of the bilayer MDs structure, and $\hat{M}_{ij}$ denotes the boundary-condition matrix relating the electric field amplitudes of the forward $(E^{+})$ and backward $(E^{-})$ waves at the interface between the layers $i$ and $j$
\begin{equation}
\left(
\begin{tabular}{ccc}
$E_{j}^{+}$ \\
$E_{j}^{-}$
\end{tabular}
\right)
=\hat{M}_{ij}
\left(
\begin{tabular}{ccc}
$E_{i}^{+}$ \\
$E_{i}^{-}$
\end{tabular}
\right),
\end{equation}
and
\begin{equation}
\hat{M}_{ij}=\frac{\mu_{j}^{2}-\Delta_{j}^{2}}{2\mu_{j}k_{xj}}
\left(
\begin{tabular}{ccc}
$f^{*}_{j}+f_{i}$ & $f^{*}_{j}-f^{*}_{i}$ \\
$f_{j}-f_{i}$ & $f_{j}+f^{*}_{i}$
\end{tabular}
\right),
\end{equation}
where $f_{m}=(\mu_{m}k_{xm}+i\Delta_{m}k_{ym})/(\mu_{m}^{2}-\Delta_{m}^{2}) (m=i,j)$ and $\hat{P}_{m}$ represents the
usual propagation matrix
\begin{equation}
\hat{P}_{m}=
\left(
\begin{tabular}{ccc}
$e^{ik_{xm}a_{m}}$ & $0$ \\
$0$ & $e^{-ik_{xm}a_{m}}$
\end{tabular}
\right).
\end{equation}
We then obtain the reflection coefficients $r_{L}$ and $r_{R}$ for the light incident from left and right:
\begin{eqnarray}
r_{L}&=&\frac{E_{0}^{-}}{E_{0}^{+}}|_{E_{3}^{-}=0}=-\frac{T_{21}}{T_{22}},\\
r_{R}&=&\frac{E_{3}^{+}}{E_{3}^{-}}|_{E_{0}^{+}=0}=\frac{T_{12}}{T_{22}}.
\end{eqnarray}
By finding the zeros of the reflectance $R_{L,R}(\equiv |r_{L,R}|^{2})$, we finally obtain the
mode solutions for the bilayer MDs structure,
\begin{eqnarray}
&&\frac{\sin (k_{x} a)}{k_{x}k_{x0}}
[\cos (k_{x} a) (k_{y}^{2}(\frac{\mu_{0}}{\mu_{1}}-\frac{\mu_{eff}}{\mu_{0}})-\frac{\omega^{2}}{c^{2}}(\epsilon_{m}\mu_{0}-\epsilon_{0}\mu_{eff})) \nonumber \\
&+&\frac{k_y \Delta_{1}}{k_x \mu_1}\sin(k_{x} a) (k_{y}^{2}(\frac{\mu_{0}}{\mu_{1}}+ \frac{\mu_{eff}}{\mu_{0}})-\frac{\omega^{2}}{c^{2}}(\epsilon_{m}\mu_{0}+\epsilon_{0}\mu_{eff}))
]   \nonumber \\
&=&0
\end{eqnarray}
where $\epsilon_{0}$ and $\mu_{0}$ are, respectively, the permittivity and permeability for surrounding medium, $k_{x0}=\sqrt{\epsilon_{0}\mu_{0}\omega^{2}/c^{2}-k_{y}^{2}}$ and $k_{x}=k_{x1}=k_{x2}=\sqrt{\epsilon_{m}\mu_{eff}\omega^{2}/c^{2}-k_{y}^{2}}$ are the wave-vector components normal to the interface in background and magnetic materials, respectively.
The solutions are then analytically separated as reciprocal (symmetrical) modes [Eq. (\ref{symm})] and non-reciprocal (asymmetrical) ones [Eq. (\ref{asymm})].

\section{APPENDIX C: Properties of scattering matrices in two- (and multi-) port systems}

We start with the scattering case in a two-port system shown in Fig. 6(a), which usually considered in the literature. It can also represent the plane-wave normal-incidence case in our paper. In this simple case, the scattering matrix equation will be in the form of
\begin{equation}
\left(
\begin{tabular}{ccc}
$E_{a}^{+}$ \\
$E_{b}^{-}$
\end{tabular}
\right)
=
\left(
\begin{tabular}{ccc}
$r_{R}$ & $t$ \\
$t$ & $r_{L}$
\end{tabular}
\right)
\left(
\begin{tabular}{ccc}
$E_{a}^{-}$ \\
$E_{b}^{+}$
\end{tabular}
\right),
\end{equation}
and the determinant of the transfer matrix and $\mathcal{PT}$ symmetry in a one-dimensional system lead to the conservation relation
$|1-T|=\sqrt{R_{L}R_{R}}$ \cite{Ge}, where $T$ is the transmittance for both sides, and $R_{L(R)}\equiv |r_{L(R)}|^{2}$ is the reflectance for wave at port $a$ ($b$). We further note that we have $R_{L}=R_{R}(=R)$ and $1-T=R$ in our plane-wave normal-incidence case since our system is ``Hermitian" and there are spatial symmetries such as  $\pi$-rotation about y-axis. In this case, there is no asymmetry in transmission although the system itself has broken reciprocity.
\begin{figure}[!htbp]
\includegraphics[width=2.2in]{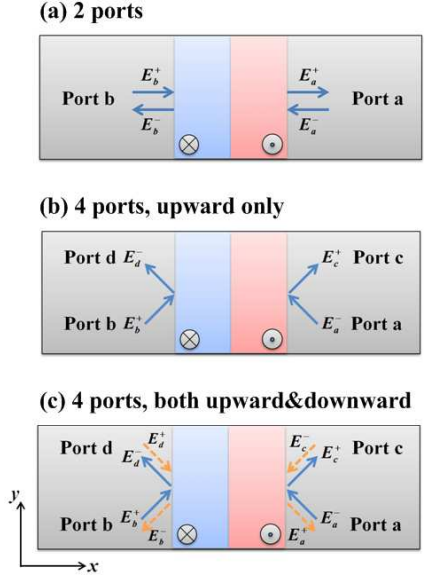}
\centering
\caption {\small (color online) (a) The usual scattering case in a two-port system. (b) ``Incomplete" off-axis scattering problem in a four-port system. (c) ``Complete" off-axis scattering problem in (b). }
\end{figure}

Figure 6(b) shows the case of ``incomplete" off-axis scattering problem in a four-port system. It can represent the ``incomplete" scattering problem in the calculation of transmittance and reflectance for a given parallel component of the wave-vector. The parallel component is directed ``upward" in Fig. 6(b). The ``complete" scattering problem will be described in Fig. (c) later. We now consider Fig. 6(b) first. The scattering matrix equation for Fig. 6(b) is in the form of
\begin{equation}
\left(
\begin{tabular}{ccc}
$E_{c}^{+}$ \\
$E_{d}^{-}$
\end{tabular}
\right)
=
\left(
\begin{tabular}{ccc}
$r_{ca}$ & $t_{cb}$ \\
$t_{da}$ & $r_{db}$
\end{tabular}
\right)
\left(
\begin{tabular}{ccc}
$E_{a}^{-}$ \\
$E_{b}^{+}$
\end{tabular}
\right),
\end{equation}
where $r_{ij}$ and $t_{ij}$ denote the reflection and transmission coefficients from port $j$ to $i$ $(i,j$ could be taken as port $a, b, c,$ or $d)$, respectively. Here $r_{db}=r_{ca}$ and $t_{da}=t_{cb}$ could be found due to the $\pi$-rotation about y-axis in our system. Mathematically, this scattering matrix equation is similar to the previous case in
Fig. 6(a) except that the ``in" ports are totally different from the ``out" ports. The conservation
equation will be the same as in case Fig. 6(a).

Figure 6(c) shows the case of ``complete" off-axis scattering in Fig. 6(b). Here, ``complete" means that it takes into account of all possible incoming and outgoing waves in all coupled ports. The scattering matrix equation for this case is in the form of
\begin{equation}
\left(
\begin{tabular}{cccc}
$E_{a}^{+}$\\
$E_{b}^{-}$\\
$E_{c}^{+}$\\
$E_{d}^{-}$\\
\end{tabular}
\right)
=
\left(
\begin{tabular}{cccc}
$0$ & $0$ & $r_{\downarrow}$ & $t_{\downarrow}$\\
$0$ & $0$ & $t_{\downarrow}$ & $r_{\downarrow}$\\
$r_{\uparrow}$ & $t_{\uparrow}$ & $0$ & $0$\\
$t_{\uparrow}$ & $r_{\uparrow}$ & $0$ & $0$\\
\end{tabular}
\right)
\left(
\begin{tabular}{cccc}
$E_{a}^{-}$\\
$E_{b}^{+}$\\
$E_{c}^{-}$\\
$E_{d}^{+}$\\
\end{tabular}
\right).
\end{equation}
Here, we use the subscripts ``$\uparrow$" and ``$\downarrow$" to denote the quantities for the ``upward" and ``downward" rays, respectively. It is also denoted by different colors in Fig. 6(c). Since the ``upward" and ``downward" modes are independent, the conservation equation can be satisfied independently, $1-T_{\uparrow}=R_{\uparrow}$ and $1-T_{\downarrow}=R_{\downarrow}$, while the scattering matrix is of the standard non-reciprocal property $S_{4\times 4}^{T}\neq S_{4\times 4}$ and thus $T_{da (or~cb)}\equiv T_{\uparrow} \neq T_{\downarrow} \equiv T_{ad (or~bc)} $ gives rise to one-way optical tunneling for oblique incidence (see Fig. 3 in our paper).

\end{document}